\documentstyle[12pt]{article}
\begin{document}
\vspace*{-1in}
\begin{flushright}
CERN-TH. 7294/94 \\
MPI-PhT/94--32\\
hep-ph/9406424 \\
\end{flushright}
\vskip 53pt
\begin{center}
{\Large{\bf Radiative B-decay as a test of CKM unitarity}}
\vskip 30pt
{\bf Gautam Bhattacharyya ${}^{\heartsuit)}$}
and {\bf Gustavo C. Branco ${}^{\clubsuit)~\star)}$}
\vskip 10pt
{\it Theory Division, CERN,\\ CH 1211, Geneva 23,  SWITZERLAND}
\vskip 10pt
and
\vskip 10pt
{\bf Debajyoti Choudhury ${}^{\diamondsuit)}$}
\vskip 10pt
{\it Max-Planck-Institut f\"{u}r Physik -- Werner Heisenberg Institut,
\\ F\"{o}hringer Ring 6, D-80805 Munich, GERMANY.}
\vskip 60pt

{\bf ABSTRACT}
\end{center}

We point out that $R \equiv Br(b\rightarrow d \gamma)/Br(b\rightarrow
s \gamma)$ is a sensitive probe of possible violation of CKM unitarity.
We compute $R$ in a minimal extension of the Standard Model containing
an additional isosinglet charge ($-1/3$) quark, which leads to a
deviation from CKM unitarity.

\vskip 45pt
\begin{flushleft}
CERN-TH. 7294/94 \\
June 1994\\
\vskip 18pt
${}^{\heartsuit)}$ gautam@cernvm.cern.ch \\
${}^{\clubsuit)}$ gbranco@cernvm.cern.ch \\
${}^{\diamondsuit)}$ debchou@iws186.mppmu.mpg.de
\end{flushleft}
\vskip 11pt
*)~ On leave of absence from:~ Departamento de Fisica and CFIF/UTL,
Instituto Superior T\'{e}cnico, Avenida Rovisco Pais, 1096 Lisboa
Cedex, Portugal.

\newpage
\setcounter{page} 1
 \pagestyle{plain}
The study of radiative $B$-decays provides an important test of the
Standard Model (SM) and is indeed a sensitive probe of physics
beyond the SM. The recent discovery, by the CLEO collaboration,
\cite{cleo} of the decay $B \rightarrow K^* \gamma$ with a branching
ratio
$Br (B\rightarrow K^* \gamma) = (4.5 \pm 1.5 \pm 0.9) \times 10^{-5}$
provides further motivation for this study.
Within the framework of the SM, the rare decays $b\rightarrow d \gamma$
and $b\rightarrow s \gamma$ provide independent measurements of
the Cabibbo-Kobayashi-Maskawa (CKM) elements $V_{td}$
and $V_{ts}$, respectively, which can be tested against the ones
measured from $B_d$-$\bar{B}_d$ and $B_s$-$\bar{B}_s$ mixings
\cite{inamilim}.

\vskip 10pt
In this note we will point out that the decay amplitudes
$b\rightarrow q \gamma$ ($q = d, s$) have a crucial dependence
 on the unitarity of the
$(3\times 3)$ CKM matrix. Thus the magnitude of
such rare decays provides
an excellent testing ground for new physics leading to the
violation of the CKM unitarity.
For definiteness, we will analyse the above radiative decays
in the framework of a minimal extension of the SM where
 a charge ($-1/3$) $SU(2)$ singlet quark is introduced.
Although we treat this extension beyond the SM
from a phenomenological viewpoint,
there are theoretically appealing
motivations behind its consideration, including $E_6$ grand-unified
theories and some  superstring-inspired
models \cite{e6}. Furthermore, isosinglet quarks provide a simple
solution \cite{bento} to the strong CP problem \cite{strcp}.
We will show that the branching fraction
$b\rightarrow d \gamma$ can change significantly due to lack
of GIM cancellation resulting from unitarity violation in the new
CKM matrix. However, the branching fraction $b\rightarrow s \gamma$ is
not affected to any level of significance. Therefore, the most
important impact of the violation of CKM unitarity would be a change
in the ratio
$R \equiv Br(b \rightarrow d \gamma)/Br(b \rightarrow s \gamma)$
which, in the SM, provides a reliable measure of
$|V_{td}/V_{ts}|$.

\vskip 10pt
Before describing the model in some detail, let us first recall the
general structure of the radiative $b$-decay and fix our notation.
To the leading order, the quark-level
branching ratio  $b\rightarrow q\gamma$ ($q = d, s$) is given
in units of the semileptonic $b$-decay branching ratio, as
\begin{equation}
 {{Br (b\rightarrow q\gamma)} \over{Br
(b\rightarrow ce\overline{\nu})}} = {{6 \alpha}\over{\pi\rho\lambda}}
\left|{{V_{tb} V_{tq}^*}\over{V_{cb}}}\right|^2
\left[\eta^{16/23} A_\gamma + {8\over3}(\eta^{14/23} - \eta^{16/23})
A_{g} + C\right]^2,
\label{brat}
\end{equation}
where $\eta \equiv \alpha_S(M_Z)/\alpha_S(m_b) = 0.548$,
 $\rho = (1-8r^2+8r^6-r^8-24r^4 \mbox{\rm ln}r)$
 with $r = m_c/m_b$, $\lambda =
1 - 1.61\;\alpha_S(m_b)/\pi$, and $C (=-0.1766)$
is a term emanating from
a complete calculation of the leading-logarithmic QCD corrections
\cite{bargui}; $V$ is the standard CKM matrix.
It may be noted that the ${m_b}^5$ dependence in
the partial decay widths of the $b$ quark cancels out in
eq. (\ref{brat}). An ${\cal{O}}(m^2_q/m^2_b)$ part in the
branching ratio has been  neglected.
We use $Br(b\rightarrow ce\overline{\nu}) = 0.107$.
$A_\gamma$ and $A_g$ are the coefficients of the effective operators
for the magnetic and chromomagnetic moment couplings for
$b \rightarrow q$ transitions \cite{inamilim} following
from
\begin{equation}
{\cal L}_{eff} = {\sqrt{{G^2_F}\over{8\pi^3}}}
 V_{tb} V^*_{tq}~\overline{q}~
\sigma^{\mu\nu} \left[\sqrt{\alpha} A_\gamma F_{\mu\nu}
+ \sqrt{\alpha_S} A_g T_a G^a_{\mu\nu}\right] (m_b P_R
+ m_q P_L)~b,
\label{heff}
\end{equation}
where $P_{R,L} = ( 1 \pm \gamma_5)/2$.
Within the framework of the SM,
the effective operator in eq. (\ref{heff}),
to lowest order,
is obtained from the `penguin' diagrams in fig. 1a.
For the sake of making this note self-contained, we give the explicit
expressions of $A_\gamma$ and $A_g$ :
\begin{equation}
 A_\gamma = x\left[{{7-5x-8x^2}\over{24(x-1)^3}} +
{{x(3x-2)}\over{4(x-1)^4}}\ln x\right]
\label{aphsm}
\end{equation}
and
\begin{equation}
A_g = x\left[{{2+5x-x^2}\over{8(x-1)^3}} -
{{3x}\over{4(x-1)^4}}\ln x\right],
\label{agsm}
\end{equation}
where $x = m_t^2/M_W^2$.
In all our discussions we shall assume that
${m_c^2\over{M_W^2}} \simeq 0$.

\vskip 10pt
It may be noted that the derivations of eqs. (\ref{aphsm}) and
(\ref{agsm}) depend crucially on the validity of CKM unitarity.
For the sake of illustration, particularly since we deal with
a model that violates this unitarity, we present the contribution
$f_i^W$ that replaces $A_\gamma$ or $A_g$ in eq. (\ref{heff})
for each individual diagram of fig. 1a:
\begin{eqnarray}
\label{fi}
f_1^W & = & {1\over 4} \bar{\xi}_1(x) +
{1\over 2}\bar{\xi}_2(x) \nonumber \\
f_2^W & = & x\left[{1\over 2}\bar{\xi}_0(x) -
{3\over 4} \bar{\xi}_1(x)
+ {1\over 4} \bar{\xi}_2(x) \right]  \nonumber \\
f_3^W & + & f_4^W  =  {1\over 4} \bar{\xi}_1 (x)   \\
f_5^W & = & Q_t \left[\xi_0(x) - {3\over 2}\xi_1(x) +
{1\over 2}\xi_2(x)\right]
\nonumber \\
f_6^W & = & {{Q_t}\over{4}} x \left[\xi_1(x) + \xi_2(x)\right],
\nonumber
\end{eqnarray}
where $Q_t$ is the charge of the top-quark and
\begin{equation}
\begin{array}{rclcl}
\xi_n(x)
   & = & \displaystyle \int_0^1 \frac{z^{n + 1} {\rm d} z}{1 + (x - 1) z}
   & = & \displaystyle \frac{-1}{(1-x)^{n+2} }
           \left[ \ln x + \sum_{k = 1}^{n + 1}
                            (-1)^{k} \pmatrix{n + 1 \cr k \cr}
                                                \frac{x^k -1 }{k}
           \right]
     \\[2ex]
\bar{\xi}_n(x) & = & \displaystyle {1\over{x}}
\xi_n\left({1\over x}\right)\ . &&
\end{array}
\label{xi}
\end{equation}
In the event of a unitary  $V^{CKM}$, GIM cancellation ensures that
we need consider only
\begin{equation}
\begin{array}{rclcl}
\displaystyle
\sum_{j=1}^6 \{f_j^W(x) - f_j^W(0)\} & = & \displaystyle A_\gamma \ ,
   \\[2ex]
\displaystyle
{1\over Q_t}\sum_{j=5}^6 \{f_j^W(x)-f_j^W(0)\} & = & \displaystyle A_g,
\end{array}
    \label{gluni}
\end{equation}
leading to the same results
as in eqs. (\ref{aphsm}) and (\ref{agsm}) respectively.

\vskip 10pt
Now we briefly describe our
model and the essential modifications from the SM.
We introduce an extra
down-type quark ($D$), whose left- and right-handed components
are both $SU(2)$ singlets. Consequently, the CKM matrix
$V^{CKM}$ is a $(3\times 4)$ one.
Without loss of generality we can assume that the up-quark mass matrix is
diagonal.
 $V^{CKM}$ then
consists of the first three lines of the ($4\times 4$) unitary matrix
$W$ which relates the left-handed components of the down-quark
weak and mass eigenstates:
\begin{equation}
\pmatrix{~ d^\circ_i \cr~ D^\circ}_L = W \pmatrix{~d_i \cr D}_L,
\end{equation}
where $i = 1,2,3$
and the weak
eigenstates are denoted by the superscript $0$. It can
readily be seen \cite{branco} that whereas
\begin{equation}
(V V^\dagger)_{i j} = \delta_{ij}, \qquad
z_{\alpha\beta} \equiv (V^\dagger V)_{\alpha\beta}
\neq \delta_{\alpha \beta}
\qquad \qquad
(V \equiv V^{CKM}).
\end{equation}
The weak gauge currents can be expressed in terms of the mass
eigenstates as
\begin{eqnarray}
J_\mu^W & = & {g\over{\sqrt{2}}} \bar{u}_{Li} V^{CKM}_{i\alpha}
 \gamma_\mu d_{L\alpha}, \nonumber \\
J_\mu^Z & = & {g\over{\cos\theta_W}} \left[t_3^u \bar{u}_{Li}
\gamma_\mu u_{Li} + z_{\alpha\beta} t_3^d \bar{d}_{L\alpha}
\gamma_\mu d_{L\beta} - \sin^2\theta_W J_\mu^{em}\right];
\end{eqnarray}
where $i = 1,2,3$; $\alpha,\beta = 1,2,3,4$ and $t_3^u~(t_3^d)
= 1/2~(-1/2)$.
The  generic symbols $z_{\alpha\beta}$ then parametrise the
tree-level flavour
dependence of the  down-sector neutral currents.

\vskip 10pt
Within the SM sector the couplings
$z_{\alpha\beta}$ (for $\alpha\neq\beta$)
are naturally suppressed
 by the ratio of the standard quark masses
to the vector-like quark masses (see ref. \cite{branco} for details)
and are strongly constrained by experiments.
Of special interest to us are the couplings $z_{bs}$ and $z_{bd}$,
which are constrained by $Br(B \rightarrow X \mu^+
\mu^-) \leq 5.0 \times 10^{-5}$ measured by the UA1 Collaboration
\cite{ua1}, leading to the bounds \cite{branco}:
\begin{equation}
\left| z_{bd}\over{V_{cb}}\right| \leq 0.029, \qquad \qquad
\left| z_{bs}\over{V_{cb}}\right| \leq 0.029.
\label{zbdm}
\end{equation}
For convenience, we write the explicit expressions of
$z_{bd}$ and $z_{bs}$ as parts of the unitarity quadrangles:
\begin{eqnarray}
\label{zbdbs}
z_{bd} & = & V_{tb} V^*_{td} + V_{cb} V^*_{cd} + V_{ub} V^*_{ud},
 \nonumber \\
z_{bs} & = & V_{tb} V^*_{ts} + V_{cb} V^*_{cs} + V_{ub} V^*_{us}.
\end{eqnarray}

\vskip 10pt
So far we have discussed only the $W$- and $Z$-mediated interactions.
The reason why we have not emphasized the Higgs contributions
is due to the fact that they depend on the Higgs
structure we choose for the specific model with an isosinglet quark.
The minimal structure is, of course, having only one Higgs doublet as
in the SM with a bare mass term for the isosinglet.
In this case, we have checked that the contributions from
the Higgs-mediated penguins are small with respect to the dominant
contribution to $b\rightarrow d \gamma$, which arises from $W$- and
$Z$-mediated penguins.

\vskip 10pt
Armed with the above information, let us now look at the diagrams
of interest to us. For simplicity, we
assume that $z_{bs} = 0$, so that the violation of CKM unitarity
is all contained in $z_{bd}$. (We also assume
that $z_{ds} = 0$
in view of the extremely
tight restriction from $K$--$\bar{K}$ mixing.)
Note that the inclusion of the isosinglet leaves  unaffected
the apparent structure
of the individual diagrams in fig. 1a. A crucial difference, however,
is brought about by the lack of GIM suppression. The flavour independent
terms no longer cancel; rather, they lead to contributions  proportional
to $z_{bd}$.
Unitarity violation also shows up through a new set of $Z$-mediated
penguins (fig. 1b), which originate as a result of tree-level flavour
mixing at the $Z$-vertex.
The contributions $f_i^Z$ from the diagrams of fig. 1b, analogous
to the $f_1^W$ of fig. 1a, are given by ($y=m_{d,b}^2/M_Z^2
\simeq 0,~y_D = m_D^2/M_Z^2$):
\begin{eqnarray}
f_1^Z + f_2^Z &=& Q_d t_3^d
\left[a_L^d\left\{4\xi_0(y)- 6\xi_1(y) + 2\xi_2(y)\right\}
- 4 a_R^d \left\{\xi_0(y) - \xi_1(y)\right\}\right],
\nonumber \\
f_3^Z &=&  Q_d (t_3^d)^2
\left[2\xi_0(y_D)- 3\xi_1(y_D) + \xi_2(y_D)\right],
\label{fiz}
\end{eqnarray}
where $a_L^d = t_3^d - Q_d \sin^2\theta_W,
~a_R^d = -Q_d \sin^2\theta_W$,
and the integrals $\xi_i$ are listed in eq. (\ref{xi}).
As a result,  the expressions of $A_\gamma$ and $A_g$ are modified
significantly to
\begin{equation}
\begin{array}{rcl}
A_\gamma & \longrightarrow & \displaystyle A_\gamma + \left({z_{bd}
\over{V_{tb} V_{td}^*}}\right)^{IS}
(c_\gamma^W + c_\gamma^Z), \\[2ex]
A_g& \longrightarrow & \displaystyle
    A_g + \left({z_{bd}\over{V_{tb} V_{td}^*}}\right)^{IS}
(c_g^W + c_g^Z).
\end{array}
\label{agprime}
\end{equation}
Above, the supercript ($IS$) refers to the CKM elements in the
presence of an isosinglet quark;
$c_{\gamma,g}^{W,Z}$ are constants arising from the
$W(Z)$-mediated photon(gluon)-penguins. These numbers
can easily be derived from eqs. (\ref{fi}),(\ref{xi}),(\ref{fiz}) to be
\begin{eqnarray}
 c_\gamma^W & \simeq & \sum_{j=1}^6 f_j^W(0) = 23/36 \nonumber,\\
 c_g^W & \simeq & {1\over {Q_t}} \sum_{j=5}^6 f_j^W(0) = 1/3 , \\
 c_\gamma^Z & \simeq & \sum_{j=1}^2 f_j^Z(0) = -13/108 \nonumber,\\
 c_g^Z & \simeq & {1\over {Q_d}}
 \sum_{j=1}^2 f_j^Z(0) = 13/36 \nonumber.
\label{c}
\end{eqnarray}
Needless to say, one should
also add the contribution of $f_3^Z$ to the above.
This obviously depends on
$m_D$ and thus introduces an additional unknown.
However, it can easily be
checked that such contributions are typically smaller and make little
quantitative impact. For example,
even for relatively large $z_{i4}$ consistent with the unitarity of
$W$, these extra contributions, which we neglect, are $(-1/60)$ to
$c_\gamma^Z$ and $(1/20)$ to $c_g^Z$ for $y_D=2$ and are
even smaller for larger
values of $y_D$.

\vskip 10pt
The next task is to determine the element $V_{td}^{IS}$.
It ought to be stressed that the
experimentally allowed range of values of the CKM matrix elements in
this scenario differ from the corresponding ones in the SM. For
example, in the SM, the element $V_{td}$ is determined by comparing the
$B_d$--$\bar{B}_d$ mixing data with its prediction driven by the
$t$-mediated box. In this scenario, the extraction of $V_{td}$ is more
complicated, though, since
$B_d$--$\bar{B}_d$ mixing receives a tree-level
contribution due to flavour-violating $Z$ couplings in the down-sector.
The bound of eq. (\ref{zbdm}) allows for this tree level contribution
to be comparable with, or even dominate,  the $t$-mediated SM box.
 In the following, we will
neglect the effect of the lack
of CKM unitarity in the evaluation of the box
diagrams, since these effects are small compared to the tree-order
$Z$-mediated ones.
For the sake of convenience we follow the notations of
ref. \cite{branco} to write:
\begin{equation}
|V_{tb} V_{td}^*|^{IS} \left|\Delta_{bd}\right|^{1/2} = F~x_d^{1/2},
\label{del1}
\end{equation}
where
\begin{equation}
F = \left[{{6\pi^2}\over{G_F^2 \eta M_W^2 M_B}}\right]^{1/2}
{1\over{\tau_B^{1/2} B_B^{1/2} f_B}}
{|\bar{E}(x)|}^{-1/2}.
\label{f}
\end{equation}
The effects of new physics are contained in
$\left|\Delta_{bd}\right|$ in eq.(\ref{del1}), which is parametrized
as:
\begin{equation}
\Delta_{bd} = 1 + r_d e^{i2\theta_{bd}},
\label{del2}
\end{equation}
with
\begin{eqnarray}
r_d & = & {1\over{\nu |\bar{E}(x)|}} \left|{{z_{bd}}\over{(V_{tb}
V_{td}^*)^{IS}}}\right|^2,
\qquad \nu = {\alpha\over{4\pi\sin^2\theta_W}},
\nonumber \\
\theta_{bd} & = &
{\mbox{\rm arg}}\left[{z_{bd}\over{(V_{tb} V_{td}^*)^{IS}}}\right]
\label{rtheta}
\end{eqnarray}
In the above expressions, the experimental inputs are given by
\cite{alilon} $x_d = 0.71 \pm 0.07$,
{}~$\tau_B = (1.54 \pm 0.03)$~ps, and
$\sqrt{B_B f_B^2}$ lies between 110 and 270 MeV;
$\bar{E}(x)$ is the standard Inami--Lim
function for the $t$-mediated box diagram and $\nu \bar{E} = -0.0065$
for $m_t = 174$ GeV. The phase $\theta_{bd}$ is an independent
parameter, which determines the orientation of $z_{bd}$
in the unitarity quadrangle and plays an important role since it
allows for different contributions of new physics even for a fixed
$\left|z_{bd}\right|$.
Since the aim of our analysis
is to focus on the departure from the SM prediction due to unitarity
violation, we fix the experimentally derived inputs in eqs.
(\ref{del1})--(\ref{rtheta}) at their central values. As a result, our
estimates of the effects of
new physics are conservative; one could always obtain an
enhanced effect by setting the experimental inputs at their extrema.

\vskip 10pt
In eq. (\ref{rtheta}) one can put $\left|V_{tb}\right|^{IS} = 1$
as a good approximation.
Solving for $\left|V_{td}\right|^{IS}$ using
eqs. (\ref{del1})--(\ref{rtheta}) for $\theta_{bd}$ in the range
$[0^\circ-180^\circ]$ for fixed values of $\left|z_{bd}\right|$
within the
allowed domain as shown in eq. (\ref{zbdm}), we obtain the allowed
values of $\left|V_{td}\right|^{IS}$, displayed in fig. 2.
For large values of $\left|z_{bd}\right|$,
$\theta_{bd}$ is constrained to a given
range in order for the solution for $|V_{td}|^{IS}$ to exist.
We present our results only for $m_t =$ 174 GeV
and remark that they are quite insensitive to its choice in
the range $m_t = 174 \pm 17$ GeV \cite{cdftop}. The region between
the horizontal lines corresponds to the SM uncertainties, taking into
account the experimental uncertainties of the various inputs.

\vskip 10pt
In fig. 3 we present the ratio $Br(b \rightarrow d \gamma)/
Br(b \rightarrow s \gamma)$ as a function of $\theta_{bd}$ for
the same values of $z_{bd}$ as in fig. 2 and putting $z_{bs}=0$.
We  find that even if one allows for
$z_{bs} \neq 0$, the branching ratio $Br(b\rightarrow s \gamma)$
does not differ significantly from its SM value. The reason is
essentially due to the fact that although $z_{bs}$ and $z_{bd}$
could be of the same order of magnitude, $\left|V_{ts}\right|$
is much larger than $\left|V_{td}\right|$. Combining the results
of the quadrangular unitarity [eq. (\ref{zbdbs})] and the information
of the CKM matrix elements, it has been shown \cite{branco} that:
\begin{equation}
\label{zbdvszbs}
\left|{{z_{bd}}\over{V_{tb} V_{td}^*}}\right| \leq 0.93;
\qquad \qquad
\left|{{z_{bs}}\over{V_{tb} V_{ts}^*}}\right| \leq 0.04.
\end{equation}
As a result of bounds in eq. (\ref{zbdvszbs}), one concludes that in the
case of $B_d$ mixing, tree-level flavour-changing $Z$-exchange
may give a dominant contribution, while in the case of $B_s$ mixing,
the $t$-mediated box diagrams are the dominant ones \cite{branco,sil}.

\vskip 10pt
It should be noted that
in the SM the ratio $Br(b\rightarrow d \gamma)/Br(b\rightarrow
s \gamma)$ is given, to a very good approximation, by
\begin{equation}
R = {Br(b \rightarrow d \gamma)\over{Br(b \rightarrow s \gamma)}}
\simeq {\left|V_{td}\right|^2 \over{\left|V_{ts}\right|^2}}.
\label{brr}
\end{equation}
The result of eq. (\ref{brr}) is quite reliable since most of the
uncertainties, such as the value of $m_b$ and the bulk of QCD
corrections, cancel out in the ratio.

\vskip 10pt
It is clear from fig. 3 that as a result of violation of CKM unitarity,
$R$ can significantly deviate from the SM prediction, for values of
$z_{bd}$ consistent with the bound of eq. (\ref{zbdm}). Note that in
fig. 3, we have indicated for the SM the regions of allowed $R$ values
within the horizontal lines taking into account the experimental
uncertainties of the various inputs \cite{alilon},
while for the present model we have taken
central values for various experimental inputs. Obviously, the effect
of new physics becomes more ``visible" once the SM prediction is
sharpened through a more precise knowledge of $V_{td}^{SM}$ and
$V_{ts}^{SM}$.

\vskip 10pt
We should, however, point out that the experimental extraction of
the ratio $R$ of eq. (\ref{brr}) is not so straightforward on account
of both QCD corrections and uncertainties
in the hadronic matrix elements.
For decays into light quarks, exclusive final states are theoretically
``cleaner''. However, in the case of $B \rightarrow X_d \gamma$, a double
Cabibbo--suppressed process, one has a situation where the $O(\alpha_s)$
corrections involving virtual $u$ and $c$ quarks are no longer
negligible \cite{aligreub}.
The partial width is thus no longer a simple function of
$|V_{td}|^2$. Apart from this, there is also the experimental
problem of measuring the above inclusive decay rate.
The exclusive modes are easier to measure experimentally, although
they are even more difficult to handle theoretically.
As an illustration, let us consider
\begin{equation}
{\Gamma (B_{u,d} \rightarrow \rho \gamma) \over
\Gamma (B_{u,d} \rightarrow K^* \gamma)} =
{\left|V_{td}\right|^2 \over
\left|V_{ts}\right|^2} \widetilde{R} \Phi_{u,d},
\end{equation}
where $\widetilde{R} = \left|F_1^{B\rightarrow \rho} (0)\right|^2 /
\left|F_1^{B\rightarrow K^*} (0)\right|^2$, the ratio of the hadronic
form factors, and $\Phi_{u,d}$ is a
phase space factor. While in the exact
$SU(3)$ limit the hadronic uncertainties
would cancel to leave $\widetilde{R} = 1$, in the real world one has to
calculate it within some model. An idea of the uncertainties involved can
be formed by looking at the results obtained from
different approaches for
$\widetilde{R} = 0.04$ (Isgur-Scora-Grinstein-Wise formalism),
$0.6$ (QCD sum rule on the
light cone) and $0.8$ (Bauer-Stech-Wirbel
formalism) \cite{alibraun,soares}.
Another correction, in this context, has to do with the
final-state interactions, which may be important for
$B \rightarrow \rho \gamma$, but not so for $B \rightarrow K^* \gamma$
\cite{soares}.

\vskip 10pt
At present, no decay of the type $B \rightarrow X_d + \gamma$
has been observed and only upper bounds have recently been
obtained by the CLEO collaboration \cite{payne} for the following
exclusive decays (90$\%$ C.L.):
\begin{equation}
\label{cleonew}
\begin{array}{rcl}
\displaystyle Br( B  \rightarrow  \rho^{-} \gamma )
    & < & \displaystyle 1.8 \times 10^{-5}, \\
\displaystyle Br( B  \rightarrow  \rho^{0} \gamma )
   & < & \displaystyle 3.1 \times 10^{-5}, \\
\displaystyle Br( B  \rightarrow  \omega \gamma )
    & < & \displaystyle 1.4 \times 10^{-5}.
\end{array}
\end{equation}
{}  From eq. (\ref{cleonew}) and within the SM, the following bound has
been extracted \cite{payne}:
\begin{equation}
\label{cleor}
\left|{{V_{td}}\over{V_{ts}}}\right| < {1\over{1.8}},
\end{equation}
which in turn implies within the SM, using eq. (\ref{brr}),
\begin{equation}
\label{cleobr}
{Br(b \rightarrow d \gamma)\over{Br(b \rightarrow s \gamma)}}
< 0.31.
\end{equation}

\vskip 10pt
It can be readily seen that the estimation of the bound of eq.
(\ref{cleobr}) also holds in our model,
and it is compatible with the
prediction of our model even for the largest value of
$\left|z_{bd}\right|$. Furthermore, it is evident that the
experimental bound in eq. (\ref{cleonew}) is still not strong
enough for the bound of eq. (\ref{cleor}) to be competitive
with the present SM limit on the same, derived from $B_d$--$\bar{B}_d$
mixing and unitarity. However, it is clear that improved data
and the eventual detection of the above exclusive decays will
provide an important constraint on the SM and have the potential
to uncover new physics.

\vskip 10pt
The presence of flavour-changing $Z$-mediated interactions has other
phenomenological consequences. Recently, it has been pointed out
\cite{nir} that in the presence of a singlet down-type quark, the SM
prediction $7\leq x_s\leq 40$ changes to $2\leq x_s\leq 50$.
This result is specially relevant if the experimental
lower bound on $x_s$ turns out to be
smaller than the lower bound predicted by the SM. It is clear that
once the experimental value of $x_s$ is known and if one allows for
$z_{bs} \neq 0$, the extraction of $\left|V_{ts}\right|^{IS}$
would be entirely analogous to the one we have described for
$\left|V_{td}\right|^{IS}$.

\vskip 10pt
Furthermore, it has been pointed out in refs. \cite{branco,sil}
that in the
presence of violation of CKM unitarity, CP asymmetries in $B$-decays
can differ significantly from those predicted in the SM. In particular,
it has been shown \cite{branco} that the sign of CP asymmetry in the
decay $B\rightarrow \Psi K_s$ may be opposite to the one
predicted by the SM, even for rather small values of $z_{bd}$.

\vskip 10pt
At this point the following comment is in order. It is clear that the
primary task for the next generation of experiments is to check whether
all data on rare $B$-decays, $B$--$\bar{B}$ mixings and CP asymmetries
can be accommodated within the SM. If this fit is not possible,
thus implying evidence for new physics, the next task would then be
to discover what the new physics is and, in particular, whether there
is any violation of CKM unitarity, and measuring the parameters $z_{bd}$
and $\theta_{bd}$.

\vskip 10pt
To conclude, the measurement of the ratio
$Br(b \rightarrow d \gamma)$ together
with the measurement of CP asymmetry in $B$-decays will provide a
crucial test of the CKM unitarity, leading either to the discovery
of unitarity violation, or to strong constraints on the parameters
$z_{bd}$ and $\theta_{bd}$.

\vskip 30pt
\noindent{\bf Acknowledgements}
\par
G.B. and G.C.B. would like to thank A. Ali for useful conversations.
G.C.B thanks the CERN Theory Division for their hospitality, and T.
Morozumi and L. Handoko
for correspondence concerning their related work
in progress. The work of G.C.B. was supported in part by Science
Project No. SCI-CT91-0729 and EC contract No. CH RX - CT93 - 0132.

 \newpage
 \noindent{\bf Figure captions}
 \par

 \vskip 50pt
 \begin{itemize}
 \item [1.]
\subitem [a]
The $W$-mediated photon-penguins in the Feynman gauge. The
corresponding gluon-penguins are realized by replacing the external
photon lines by gluon ones in diagrams 5 and 6.

\subitem [b]
The $Z$-mediated photon-penguins in the
Feynman gauge. The longitudinal mode contributions are not
shown, as they are negligible. The
corresponding gluon-penguins are realized by replacing the external
photon lines by gluon ones in all the diagrams.

\item[2.]
Variation of $\left|V_{td}\right|^{IS}$ with $\theta_{bd}$ for different
values of $\left|z_{bd}\right|$ [$1.16\times 10^{-3}$ (dot-dashed),
$0.78\times 10^{-3}$ (dotted) and $0.39\times 10^{-3}$ (dashed)],
evaluated using the central values of
the experimental inputs ($x_d$, $\tau_B$, $\sqrt{B_B f_B^2}$ etc.)
and for $m_t = 174$ GeV. The region between the horizontal lines
corresponds to the allowed values within the SM, taking into account
the experimental uncertainties of the above inputs.

\item[3.]
Variation of $R = Br(b\rightarrow d \gamma)/Br(b\rightarrow s \gamma)$
with $\theta_{bd}$ for the same values of $z_{bd}$ and the same
central values of the experimental inputs as in fig. 2.
The horizontal band
corresponds to the allowed values of $R$ within the SM for the
uncertainties in various inputs as described in fig. 2.

\end{itemize}


\newpage

\begin{thebibliography}{99}

\bibitem{cleo} CLEO Collaboration, R. Ammar et al.,
{\it Phys. Rev. Lett.} {\bf 71} (1993) 674; \\
T. Browder, K. Honscheid
and  S. Playfer, in ``{\it B Decays}", 2nd edition, ed. S. Stone
(World Scientific, Singapore, 1994).


\bibitem{inamilim} T. Inami and C.S. Lim, {\it Prog. Theor. Phys.}
{\bf 65} (1981) 297.


\bibitem{e6} F. Zwirner, {\it Int. J. Mod. Phys.} {\bf A3} (1988) 49;
\\ J.L. Hewett and T.G. Rizzo,
{\it Phys. Rep.} {\bf 183} (1989) 193 and references therein.

\bibitem{bento} L. Bento, G.C. Branco and P.A. Parada, {\it Phys.
Lett.} {\bf B267} (1991) 95.

\bibitem{strcp} A. Nelson, {\it Phys. Lett.} {\bf B136} (1984) 387; \\
S.M. Barr, {\it Phys. Rev. Lett.} {\bf 53} (1984) 329.

\bibitem{bargui}
R. Barbieri and G. Giudice, {\it Phys. Lett.} {\bf B309} (1993) 86;\\
M. Ciuchini et al., {\it Phys. Lett.} {\bf B316} (1993) 127;\\
M. Misiak, {\it Phys. Lett.} {\bf B269} (1991) 161.

\bibitem{branco} G. Branco, T. Morozumi, P.A. Parada and M.N. Rebelo,
{\it Phys. Rev.} {\bf D48} (1993) 1167.

\bibitem{ua1} UA1 Collaboration,
C. Albajar et al., {\it Phys. Lett.} {\bf B262} (1991) 163.

\bibitem{alilon} See e.g. A. Ali and D. London,
preprint CERN-TH. 7248/94 (1994).

\bibitem{cdftop} CDF Collaboration, F. Abe et al.,
preprint FERMILAB-PUB-94/097-E (1994).

\bibitem{sil} Y. Nir and D. Silverman, {\it Phys. Rev.} {\bf D22}
(1990) 1477; \\ D. Silverman, {\it Phys. Rev.} {\bf D45} (1992) 1800;\\
A. Ray-Mukhopadhyaya and A. Raychaudhuri, {\it Phys. Rev.} {\bf D37}
(1988) 807; \\ B. Mukhopadhyaya, A. Ray and A. Raychaudhuri,
{\it Phys. Lett.} {\bf B186} (1987) 147.

\bibitem{aligreub} A. Ali and C. Greub, {\it Phys. Lett.}
{\bf B287} (1992) 191.

\bibitem{alibraun} A. Ali, V.M. Braun and H. Simma, preprint
CERN-TH.7118/93, MPI-Ph/93-77, DESY 93-193.

\bibitem{soares} J.M. Soares, {\it Phys. Rev.} {\bf D49} (1994) 283.

\bibitem{payne} D. Payne, talk representing the
CLEO collaboration, Beauty '94, April 1994.

\bibitem{nir} Y. Nir, {\it Phys. Lett.} {\bf B327} (1994) 85.

\end{thebibliography}
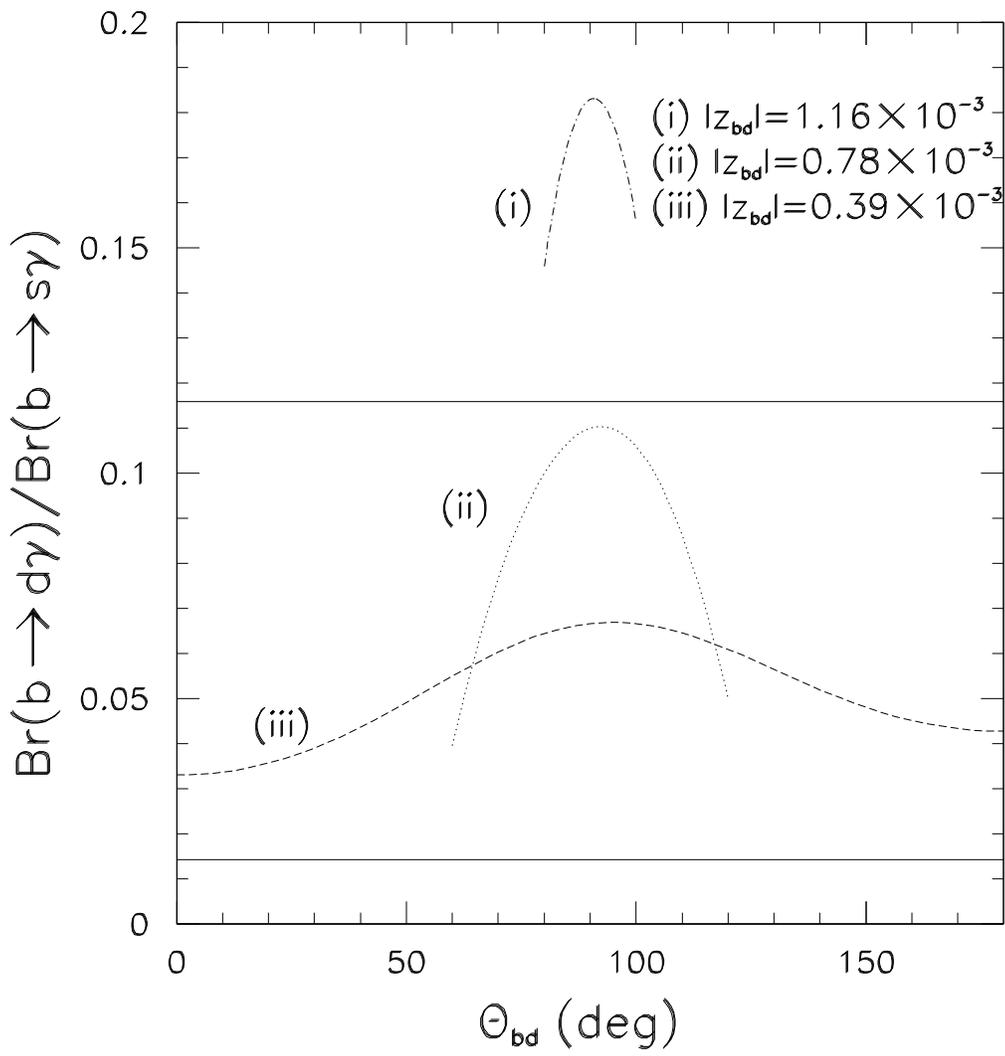
\end{document}